\documentclass{elsart}

\usepackage{graphicx}
\usepackage{amssymb}
\usepackage{epsfig}
\usepackage{subfigure}
\usepackage{color}
\usepackage[english]{babel}
\usepackage{latexsym}
\usepackage{amsfonts}
\usepackage{amsmath}
\usepackage{multirow}
\usepackage{float}

\def\be{\begin{equation}}
\def\ee{\end{equation}}

\begin{document}
\begin{frontmatter}

\title{Thermal hadron production in relativistic nuclear collisions: 
the hadron mass spectrum, the horn, and the QCD phase transition}

\author[gsi]{A.~Andronic},
\author[gsi,tud]{P.~Braun-Munzinger},
\author[hei]{J.~Stachel}

\address[gsi]{EMMI, GSI Helmholtzzentrum f\"ur Schwerionenforschung,
D-64291 Darmstadt, Germany}
\address[tud]{Technical University Darmstadt, D-64289 Darmstadt, Germany}
\address[hei]{Physikalisches Institut der Universit\"at Heidelberg,
D-69120 Heidelberg, Germany}

\begin{abstract}
  We present, using the statistical model, a new analysis of hadron
  production in central collisions of heavy nuclei. This study is motivated by the
  availability of final measurements both for the SPS (beam energies 20-160
  AGeV) and for the RHIC energies ($\sqrt{s_{NN}}$=130 and 200 GeV) and by
  updates in the hadron mass spectrum, which is a crucial input for statistical
  models.  Extending previous studies by inclusion of very high-mass
  resonances (m$>$ 2 GeV), and the up-to-now neglected scalar $\sigma$ meson
  leads to an improved description of the data. In particular,  the
  hitherto poorly reproduced energy dependence of the $K^+/\pi^+$ ratio at SPS
  energies (``the horn'') is now well described through the connection to
  the hadronic mass spectrum and, implicitly, Hagedorn's limiting temperature. 
  We thereby demonstrate the intimate connection between the horn and the QCD 
  phase transition. 
\end{abstract}

\end{frontmatter}

\section{Introduction}

One of the major goals of ultrarelativistic nuclear collision studies is to
obtain information on the QCD phase diagram \cite{pbm_wambach}.  A promising
approach is the investigation of hadron production.  Hadron yields measured in
central heavy ion collisions from AGS up to RHIC energies can be described very
well 
\cite{agssps,satz,heppe,cley,beca1,rhic,nu,beca2,rapp,becgaz,aa05,man08,letessier05}
within a hadro-chemical equilibrium model.  In our approach
\cite{agssps,heppe,rhic,aa05} the only parameters are the chemical freeze-out
temperature $T$ and the baryo-chemical potential $\mu_b$ (and the fireball
volume $V$, in case yields rather than ratios of yields are fitted); for a
review see \cite{review}.

The main result of these investigations was that the extracted temperature
values rise rather sharply from low energies on towards $\sqrt{s_{NN}}\simeq$10 GeV
and reach afterwards constant values near $T$=160 MeV, while the baryochemical
potential  decreases smoothly as a function of energy.
The limiting temperature \cite{hagedorn85} behavior suggests a connection to 
the phase boundary and it was, indeed, argued \cite{wetterich} that the 
quark-hadron phase transition drives the equilibration dynamically, at least 
for  SPS energies and above. 
Considering also the results obtained for elementary collisions, where similar
analyses of hadron multiplicities, albeit with several additional, non-statistical 
parameters (see \cite{aa08,becattini08} and refs. therein),  yield 
also temperature values in the range of 160 MeV, alternative
interpretations were put forward. These include conjectures  that the thermodynamical state  
is not reached by dynamical equilibration among constituents but rather 
is a generic fingerprint of hadronization \cite{stock,heinz}, or is a feature 
of the excited QCD vacuum \cite{castorina}. The analysis presented below lends
further support to the interpretation that the phase boundary is reflected in
features of the hadron yields. 

While in general all hadron yields are described rather quantitatively, a
notable exception was up-to-now the energy dependence of the $K^+/\pi^+$ ratio which
exhibits a rather marked maximum, ``the horn'', near $\sqrt{s_{NN}}\simeq$ 10
  GeV \cite{na49pi}. The existence 
of such a maximum was, in fact, predicted \cite{pbm4} within the framework of the
statistical model, but the observed rather sharp structure could not be
reproduced \cite{aa05}. 
Other attempts to describe the energy dependence of the $K^+/\pi^+$ ratio within 
the thermal model \cite{becgaz,letessier05} were also not successful, except 
when an energy-dependent light quark fugacity was used as an additional parameter
\cite{letessier05}.
Furthermore, all attempts to reproduce this structure
within the framework of hadronic cascade models also failed, as is discussed 
in detail in \cite{na49pi}. As a consequence, the horn structure is taken in
\cite{na49pi} as experimental evidence for the onset of deconfinement and
quark-gluon plasma formation, and as support for the predictions of \cite{gaz}.

In this letter, we present a new analysis of hadron production in central
nucleus-nucleus collisions. The motivation for this is twofold: 
i) on the
experimental side the data set has been recently extended (and/or finalized),
in particular for the SPS energy regime by the NA49 collaboration
\cite{na49pi,na49hyp,na49phi} and for the RHIC energies by the STAR
collaboration \cite{star08}, and 
ii) we would like to explore the consequences
of an improved hadronic mass spectrum in which the $\sigma$ meson and many
higher-lying resonances are included.
We note that, with the exception of the $\sigma$ meson, the full hadronic mass
spectrum has already been used in our recent investigation \cite{aa08} of hadron
production in $e^+e^-$ collisions.
In the following we first discuss the update in the hadronic mass spectrum and
then explore its consequences for the description of all available data from
SIS to RHIC energies. 

For the hadronic mass spectrum we are using the complete mass states published
recently by the Particle Data Group (PDG) \cite{pdg}.  Not all the 
states have known branching ratios and in such cases we have assigned
values based on analogies to the nearest states with the same quantum
numbers.  The systematic uncertainty introduced by this is estimated to be
well below the bias introduced if the high-mass states would not be
considered.  Another recent development in the field of hadron spectroscopy is
the strengthening of the case for the existence of the $\sigma$ meson
(labelled $f_0(600)$ in the PDG compilation \cite{pdg}) \cite{gar07,cap06,bon07}. 
The $\sigma$ meson is
a broad structure, whose properties are extracted from fits of measurements in
various channels (see \cite{gar07} for a recent review) and decays into 
$\pi^+\pi^-$.  We have adopted as ``nominal'' the values for the mass and
(Breit-Wigner) width\footnote{We ignore here that the width of the $\sigma$
  meson is not small compared to its mass. Further investigations will have to
  deal with this issue but we note that for the $\rho$ meson the situation is
  not so different.} as extracted in ref.~\cite{gar07}, $m_{\sigma}$=484 MeV,
$\Gamma_{\sigma}$=510 MeV and will investigate the effect of different
parameters, namely also the case $m_{\sigma}$=600 MeV,
$\Gamma_{\sigma}$=600 MeV.

\begin{figure}[hbt]
\centering\includegraphics[width=.65\textwidth]{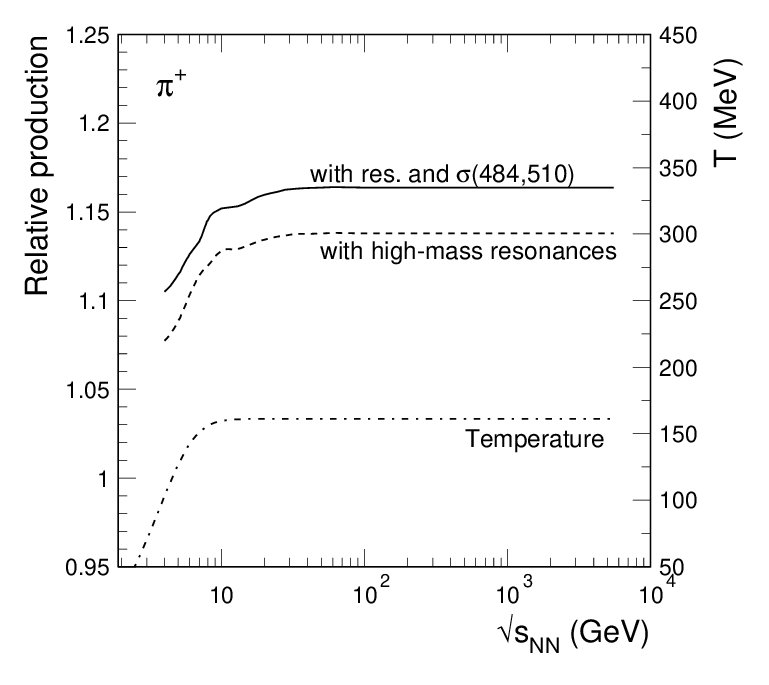}
\caption{The energy dependence of the increase in pion yield after 
inclusion of high-mass resonances (dashed) and the $\sigma$ meson,
characterized by its mass and width ($m_{\sigma}$,$\Gamma_{\sigma}$) in MeV.
The dash-dotted line depicts the energy dependence of the temperature \cite{aa05} 
used in the calculations.}
\label{fig1}
\end{figure}

To explore the consequences of the improved mass spectrum we show as a
function of energy, in 
Fig.~\ref{fig1},  the increase of yields  
of pions after inclusion of high-mass resonances, relative to the case when 
hadrons up to a mass of 2 GeV were considered, as in our earlier study 
\cite{aa05}, and after the inclusion of the $\sigma$ meson. The observed
energy dependence is driven mostly by the change in temperature, also shown 
in Fig.~\ref{fig1}.
The parametrizations for the energy dependence of $T$ and $\mu_b$ established in
\cite{aa05} were employed. The high-mass resonances lead to an increase of 
about 13\% for the calculated pion yields. This increase levels off 
near the point where the temperature reaches its limiting value,
thereby sharpening the structure in the $K^+/\pi^+$ ratio, as will be
demonstrated below.
A further 3.5\% increase of the calculated pion yield is due to the presence 
of the $\sigma$ meson, with a rather small dependence on its mass and width.
For $m_{\sigma}$=600 MeV, $\Gamma_{\sigma}$=600 MeV the calculations lead to
about 1\% fewer pions.
For the $\Lambda$ hyperons, the new high mass resonances lead to an increase 
in the calculated production of about 22\%, while the addition of the $\sigma$ 
meson has no effect.
An increase of up to 6\% is observed for protons, while for kaons this increase 
is about 7\%.

We have shown earlier \cite{aa05} that the thermal fits of hadron yields 
and of ratios of yields lead to very similar results. For the present analysis
we focus on fits of yields.
We mostly utilize mid-rapidity data, but, at lower SPS energies, fit also
the hadron yields for the full phase space. 

\begin{figure}[hbt]
\begin{tabular}{lr} \begin{minipage}{.49\textwidth}
\centering\includegraphics[width=1.\textwidth]{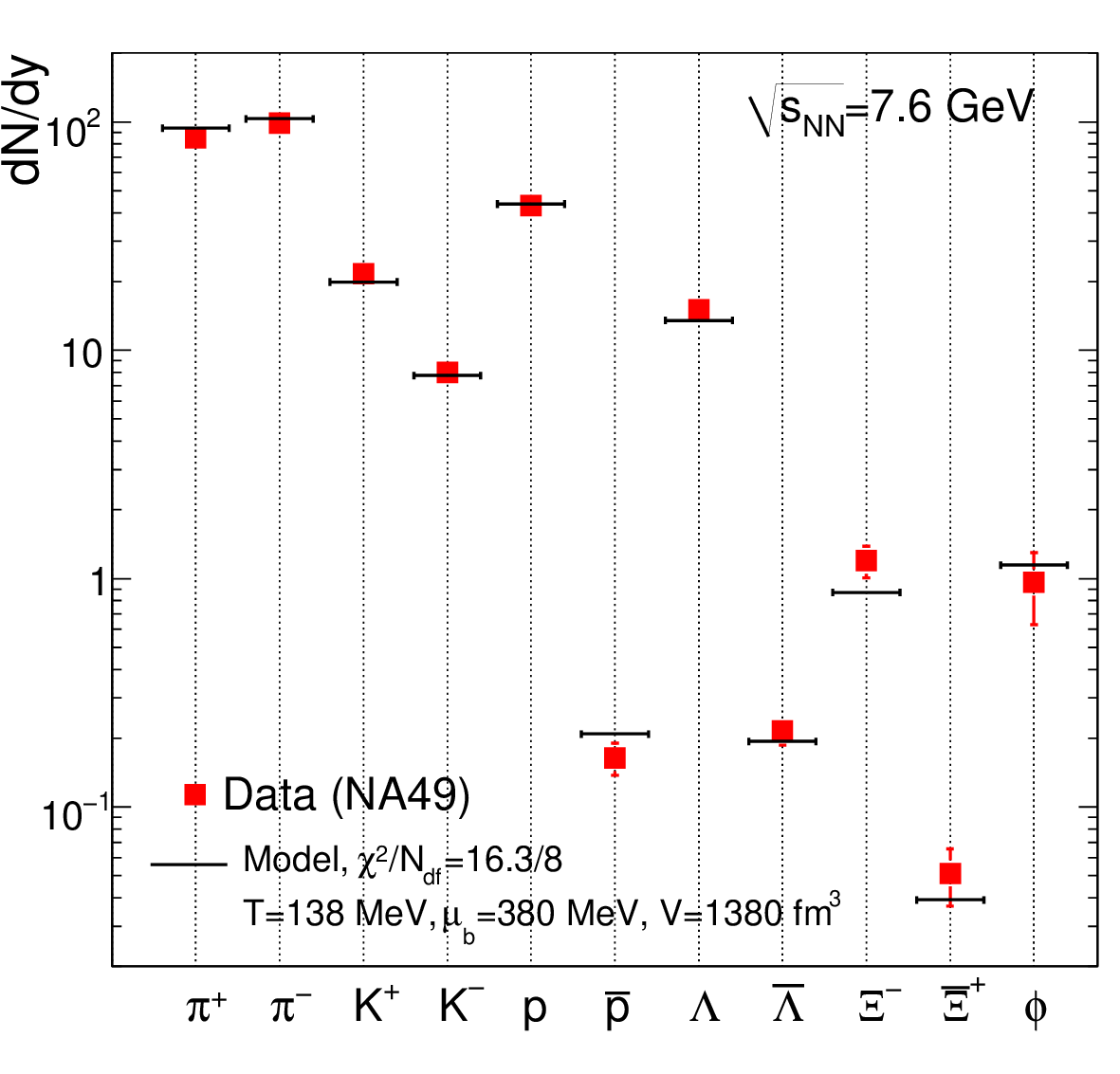}
\end{minipage} &\begin{minipage}{.49\textwidth}
\centering\includegraphics[width=1.\textwidth]{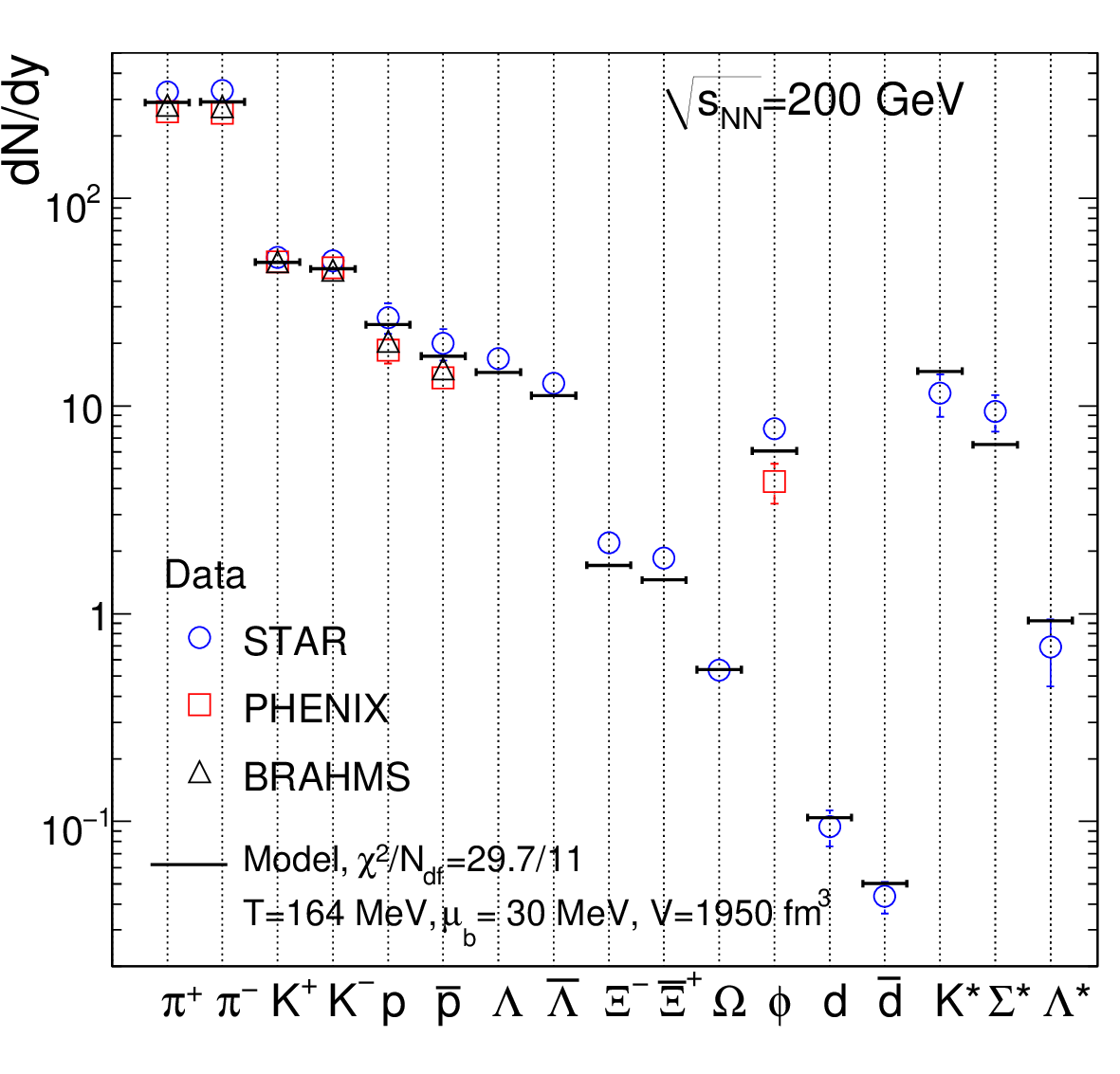}
\end{minipage} 
\end{tabular}
\caption{Experimental hadron yields and model calculations for the parameters 
of the best fit at the energies of 7.6 (left panel) and 200 GeV (right panel;
the $\Omega$ yield includes both $\Omega^-$ and $\bar{\Omega}^+$).}
\label{fig_fits}
\end{figure}

In Fig.~\ref{fig_fits} we present a comparison of measured and calculated
hadron yields at the energies of  
$\sqrt{s_{NN}}$=7.6 GeV (beam energy of 30 AGeV at SPS) and 
$\sqrt{s_{NN}}$=200 GeV. The model is successful in reproducing 
the measurements and this applies to all energies, from 2 AGeV beam energy
(fixed target) up to the top RHIC energy of $\sqrt{s_{NN}}$=200 GeV. 
The reduced $\chi^2$ values are reasonable.
In most cases the fit quality is improved compared to
our earlier analysis \cite{aa05}, even though the experimental errors are 
now smaller.
Whenever several independent measurements are available, we have employed 
a weighted mean of the data following the recipe given in the introduction
of \cite{pdg}.
A special case is that of the top SPS energy ($\sqrt{s_{NN}}$=17.3 GeV), 
where the disagreement between the NA49 and the NA57 data persists.
As in the case of our earlier analysis \cite{aa05}, we have moved the
difference in the fit results into the respective systematic error.
A disagreement between the experiments is seen at the top RHIC energy for 
pions and protons, see Fig.~\ref{fig_fits}, which is the reason of the 
large reduced $\chi^2$. A fit of ratios is in this case more suited,
but we note that a fit of the STAR yields alone gives $T$=162 MeV, $\mu_b$=32 MeV, 
$V$=2400 fm$^3$, with a very good $\chi^2/N_{df}$=9.0/11. The resonances were 
not included in the fits, but are quite well reproduced by the model.

\begin{figure}[hbt]
\centering\includegraphics[width=.7\textwidth]{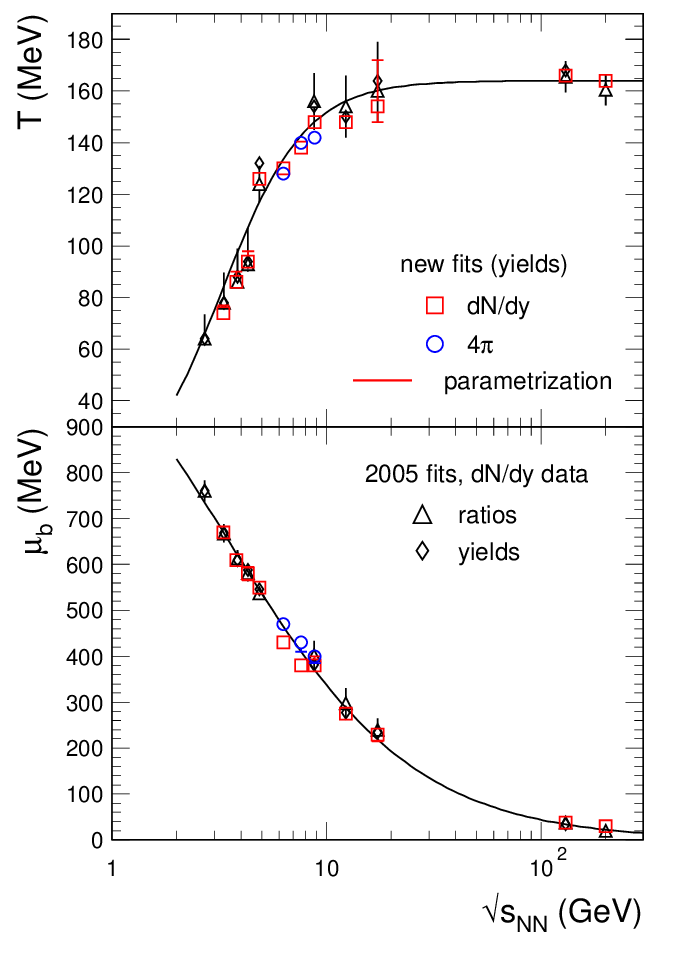}
\caption{The energy dependence of temperature and baryon chemical potential at
  chemical freeze-out.
The  results obtained here are compared to the values obtained in our earlier study 
\cite{aa05}.
The lines are parametrizations for $T$ and $\mu_b$ (see text).}
\label{fig_tmu}
\end{figure}

An important result of our analysis is that the resulting thermal parameters
are close to those obtained earlier \cite{aa05} and are in agreement 
with other recent studies \cite{man08,star08}.  
This indeed confirms that the common practice of including in the thermal codes 
hadrons up to masses of 2 GeV (for instance in the publicly-available code 
THERMUS \cite{thermus}) does not lead to significantly biased fit parameters.
Nevertheless, there are small variations.
In Fig.~\ref{fig_tmu} we present the energy dependence of $T$ and $\mu_b$ 
in comparison to our earlier results \cite{aa05}.
We have parametrized $T$ as a function 
of $\sqrt{s_{NN}}$ with the following expression\footnote{For $\mu_b$, 
  there is no need to change our earlier\cite{aa05} parametrization:  
$\mu_b \mathrm{[MeV]}=\frac{1303}{1+0.286\sqrt{s_{NN}(\mathrm{GeV})}}$}:
\be
T \mathrm{}=T_{lim}\frac{1}{1+\exp(2.60-\ln(\sqrt{s_{NN}(\mathrm{GeV})})/0.45)},
\label{pt}
\ee
with the "limiting" temperature $T_{lim}$=164 MeV. This value is slightly 
higher compared to our earlier value of 161$\pm$4 MeV \cite{aa05} due to 
the higher temperatures presently derived for the RHIC energies.
The approach to $T_{lim}$ is presently more gradual compared to our earlier
parametrization.

The values of $\mu_b$ extracted for the two lowest SPS energies deviate
somewhat from the continuous trend suggested by all the other points.  At these
energies the fit to data in full phase space does lead, as expected, to larger
values of $\mu_b$, which do fit in the systematics. At 40 AGeV
($\sqrt{s_{NN}}$=8.8 GeV) the resulting values of $T$ and $\mu_b$ from the fit
of 4$\pi$ data are very similar to those obtained from midrapidity data.

\begin{figure}[htb]
\centering\includegraphics[width=.7\textwidth]{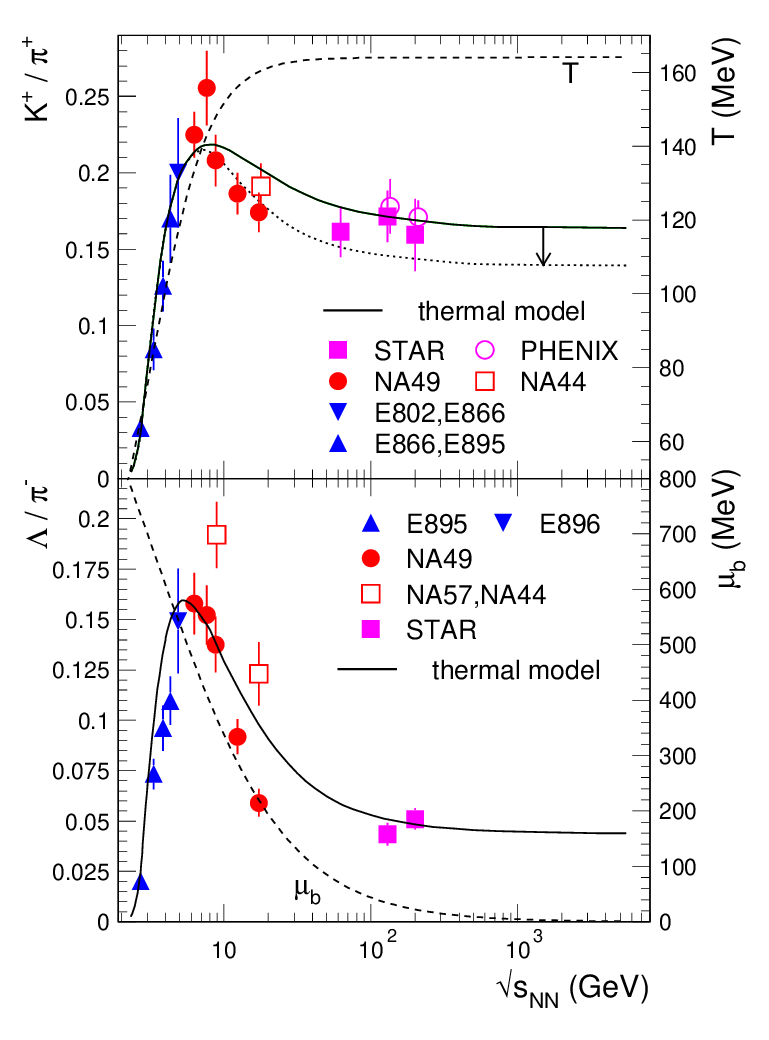}
\caption{Energy dependence of the relative production ratios $K^+/\pi^+$ 
and $\Lambda/\pi^-$. 
With the dotted line we show for the $K^+/\pi^+$ ratio an estimate
of the effect of higher mass resonances (see text).
The dashed lines show the energy dependence of $T$ (upper panel) and 
$\mu_b$ (lower panel).}
\label{fig_k2pi}
\end{figure}

We employ the above parametrization to investigate the energy dependence
of the relative production yields $K^+/\pi^+$ and $\Lambda/\pi^-$,
shown in Fig.~\ref{fig_k2pi}.
The $K^+/\pi^+$ ratio shows a rather pronounced maximum at a beam 
energy of 30 AGeV \cite{na49pi}, and the data are well reproduced by the model
calculations.
In the thermal model this maximum occurs naturally at 
$\sqrt{s_{NN}}\simeq$8 GeV \cite{pbm4}. It is due to the counteracting effects
of the steep rise and saturation of $T$ and the strong monotonous decrease 
in $\mu_b$.
The competing effects are most prominently reflected in the energy dependence
of the $\Lambda$ hyperon to pion ratio (lower panel of Fig.~\ref{fig_k2pi}), 
which shows a pronounced maximum at $\sqrt{s_{NN}}\simeq$5 GeV. 
This is reflected in the $K^+/\pi^+$ ratio somewhat less directly; it appears 
mainly as a consequence of strangeness neutrality\footnote{Recent studies within 
the UrQMD model \cite{memo} suggest the possible presence of net strangeness 
at midrapidity; the present analysis lends no support to this conjecture.},
assumed in our calculations.

The model describes the $K^+/\pi^+$ data very well\footnote{On a statistical basis,
with a relatively good $\chi^2/N_{df}$=21/12.} over the full energy range,
as a consequence of the inclusion in the code of the high-mass resonances and 
of the $\sigma$ meson, while our earlier calculations \cite{aa05} were 
overpredicting the SPS data. At RHIC energies, the quality of the present fits 
is essentially unchanged compared to \cite{aa05}, as also the data have changed
somewhat.
The model also describes accurately the $\Lambda/\pi^-$ data.
We note that the maxima in the two production ratios are located at 
different energies \cite{cleymans04}. The model calculations reproduce this 
feature in detail.

The calculated $K^+/\pi^+$ ratio (solid line in Fig.~\ref{fig_k2pi}) is likely 
to decrease further at energies beyond the maximum and the peak is likely 
to sharpen somewhat if our presumably still incomplete knowledge of the hadronic
spectrum for masses larger than 2 GeV would improve.
To get a qualitative idea about the size of the effect we have assumed that 
the true mass spectrum rises exponentially, with a Hagedorn temperature 
parameter $T_H \approx$ 200 MeV (as observed in the mass range up to 2 GeV). 
Based on this spectrum we have calculated the modification of the $K^+/\pi^+$ 
ratio by inclusion and exclusion of states above 3 GeV. The strongest 
contributions from these states to kaons comes from the decay of $K^*$ mesons 
(where only 1 kaon is produced for each additional $K^*$) while all high mass 
resonances produce multiple pions, thereby reducing the $K^+/\pi^+$ ratio. 
Using these assumptions and  Boltzmann suppression of high masses we arrive 
at the dotted line in Fig.~\ref{fig_k2pi}  While the qualitative nature of the 
estimate should be underlined, it shows in any case the expected features, 
namely a sharpening of the peak and a further reduction in $K^+/\pi^+$ at high 
energies.
The uncertainty of the calculations due to the mass and width of the $\sigma$ 
meson are at the percent level only. 
Another few percent uncertainty, which is difficult to
assess quantitatively, arises from the unknown branching ratios of the 
high-mass resonances.

\begin{table}[hbt]
\caption{Predictions of the thermal model for selected hadron ratios in central 
Pb+Pb collisions at LHC.}
\label{tab3}
\begin{tabular}{cccccc}
$p/\pi^+$ & $K^+/\pi^+$ & $K^-/\pi^-$ & $\Lambda/\pi^-$ & 
$\Xi^-/\pi^-$ & $\Omega^-/\pi^-$ \\ \hline  
0.072 & 0.164 & 0.163 & 0.042 & 0.0054 & 0.00093 \\ 
\end{tabular}
\end{table}

In Table~\ref{tab3} we present our updated predictions for the LHC energy
($\sqrt{s_{NN}}$=5.5 TeV) \cite{aa_lhc}. 
%The predicted ratios are smaller by up to 12\% ($\Lambda/\pi$ larger by 7%)
%compared to our previous results \cite{aa_lhc}.

In summary, we have demonstrated that by inclusion of the $\sigma$ meson and
many higher mass resonances into the resonance spectrum employed in the
statistical model calculations an improved description is obtained of hadron
production in central nucleus-nucleus collisions at ultra-relativistic
energies. The most dramatic improvement is visible for the $K^+/\pi^+$ ratio,
which is now well described at all energies. The ``horn'' finds herewith a
natural explanation which is, however, deeply rooted in and connected to 
detailed features of the hadronic mass spectrum which leads to a limiting 
temperature and contains the QCD phase transition \cite{hagedorn85}. 
It is interesting to note that central questions in hadron spectroscopy such 
as the existence (and nature) of the $\sigma$ meson apparently play an important 
role in quark-gluon plasma physics. 
Our results strongly imply that hadronic observables near and above the horn 
structure at a beam energy of 30 AGeV provide a link
to the QCD phase transition. Open questions are whether the chemical
freeze-out curve below the horn energy actually traces the QCD phase boundary
at large values of chemical potential or whether chemical freeze-out in this
energy range is influenced by exotic new phases such as have been predicted in
\cite{mclerran_pisarski}. In any case these are exciting prospects for new
physics to be explored at RHIC low energy runs and, in particular, at the high
luminosity FAIR facility. It remains a challenge, though, to identify clear
signals for quark-gluon plasma properties in this energy range.
At the high energy frontier, the measurements at LHC will be a crucial test 
of the present picture.

Acknowledgements:   The authors would like to thank Jochen Wambach for
pointing out the importance and special role of the $\sigma$ meson in 
the hadronic mass spectrum. 
We acknowledge the support from the Alliance Program of the 
Helmholtz-Gemeinschaft.

\end{document}